\documentclass[manuscript, screen]{acmart}

\AtBeginDocument{%
  }

\setcopyright{acmlicensed}
\copyrightyear{2025}
\acmYear{2025}
\acmDOI{XXXXXXX.XXXXXXX}

\usepackage{tikz}
\usepackage{subcaption}
\newcommand{\VR}{\textit{VR }}
\newcommand{\RW}{\textit{RW }}
\newcommand{\AVR}{\textit{AVR }}
\begin{document}

\title{From Brick to Click: Comparing LEGO Building in Virtual Reality and the Physical World}

\author{Viktorija Paneva}
\email{viktorija.paneva@ifi.lmu.de}
\orcid{https://orcid.org/0000-0002-5152-3077}
\affiliation{%
  \institution{LMU Munich}
  \city{Munich}
  \country{Germany}
}
\affiliation{%
  \institution{University of the Bundeswehr Munich}
  \city{Munich}
  \country{Germany}
}

\author{Maximilian David}
\email{david.maximilian@yahoo.de}
\orcid{https://orcid.org/0009-0009-4053-6487}
\affiliation{%
  \institution{University of Bayreuth}
  \city{Bayreuth}
  \country{Germany}
}

\author{Jörg Müller}
\email{joerg.mueller@uni-bayreuth.de}
\orcid{https://orcid.org/0000-0002-4971-9126}
\affiliation{%
  \institution{University of Bayreuth}
  \city{Bayreuth}
  \country{Germany}
}
\renewcommand{\shortauthors}{}

\begin{abstract}
We present a comparative study of building with LEGO in three environments: the physical world, a Virtual Reality (VR) counterpart, and a VR setting enhanced with "superpowers".
The study aims to understand how traditional creative hands-on activities translate to virtual environments, with potential benefits for educational, training, entertainment, and therapeutic uses.
22 participants engaged in both structured assembly and creative free-building tasks across these environments. 
We investigated differences in user performance, engagement, and creativity, with a focus on how the additional VR functionalities influenced the building experience. 
The findings reveal that while the physical environment offers a familiar tactile experience, VR, particularly with added superpowers, was clearly favoured by participants in the creative free-building scenario. 
Our recommendations for VR design include balancing automation with user control to enhance task efficiency while maintaining engagement, and implementing intuitive systems that manage complexity to prevent user overwhelm and support creative freedom.

\end{abstract}


\begin{teaserfigure}
  \includegraphics[width=\textwidth]{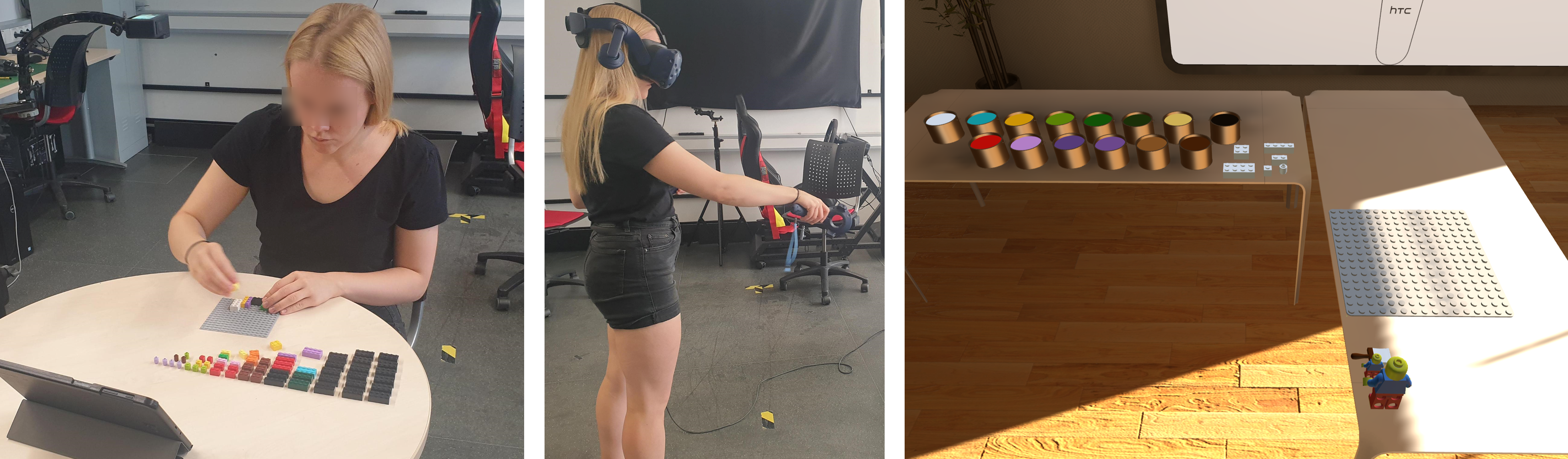}
  \caption{We compare the experience of building with LEGO across three environments: (A) the physical world, (B) Virtual Reality (VR), and (C) VR with "superpowers" (unlimited brick supply, colourable bricks, automatic brick provision, and the ability to shrink to a minifigure's perspective and explore a structure from within) .
  The study investigates differences in user performance and engagement, as well as the impact of the VR superpowers on creativity and enjoyment.}
  \Description{}
  \label{fig:teaser}
\end{teaserfigure}


\maketitle

\section{Introduction}

Playing and constructing with building blocks is experienced as creative and enjoyable by both children and adults, with prior research suggesting positive impact on learning, creativity, and the development of social skills~\cite{Wolfgang2003, freeman2003teaching, Legoff2006, Nath2014}.
The integration of constructive play into Virtual Reality (VR) presents new possibilities for location-independent, flexible and personalised creative expression and skill development. 
Engaging in activities that promote learning and creativity in VR, could, in turn, positively impact the overall VR experience, enhancing immersion, user engagement, and enjoyment. 
However, a significant challenge remains: limited dexterity and tactile feedback of currently available VR systems, as developing realistic haptic interfaces for VR remains a challenge~\cite{wee2021haptic}. 
This limitation could lead to a less authentic building experience compared to the physical world.
Nonetheless, the digital environment also offers unique capabilities that are difficult, if not impossible, to replicate in the physical world.

To fully harness its potential, it is crucial to understand how the experience of building with blocks - specifically, we use LEGO\footnote{www.lego.com} in our study, as the most prominent example - transfers from the physical to the virtual realm. 
This entails understanding the trade-offs between the limitations imposed by current VR technologies and the additional functionalities that VR can provide. 
Key questions arise: How do users perceive performing a familiar physical activity, such as building with LEGO, in a virtual environment? How does their performance in VR compare to the physical world? What new features would users like to have to enhance their virtual building experience?

To address these questions, we developed a VR application that simulates LEGO construction and conducted a user study to compare user experience and performance across three environments: the physical world (\autoref{fig:teaser}A), a VR counterpart (\autoref{fig:teaser}B), and VR with additional features, referred to as \textit{superpowers} (\autoref{fig:teaser}C).
Our study aims to answer two main questions: 
\begin{enumerate}
    \item How does the user experience and performance differ between LEGO building in the physical world and its immediate virtual counterpart?
    \item What effect do the virtual superpowers have on user engagement and creativity?
\end{enumerate}
By exploring these aspects, we aim to inform the design of future VR applications that incorporate building block activities, with potential implications for education, training, and therapeutic settings.





\section{Related Work}
In this section, we review previous research on the transferability of skills between physical and virtual environments across different tasks and applications, and the use of virtual building blocks to enhance learning experiences and foster creativity in virtual and mixed reality environments.

\textit{Transferability of Skills between Physical and Virtual Environments.}
The transferability of skills between physical and virtual environments has been a subject of great interest in prior research. 
This encompasses a diverse range of tasks, including surgical and dental procedures~\cite{Ayoub2019application}, driving~\cite{Risto2014}, climbing~\cite{Kosmalla2017}, throwing~\cite{tirp2015virtual, Zindulka2020, Kelling2023}, pipetting~\cite{petersen2022pipetting}, table tennis~\cite{Oagaz2022}, pump maintenance~\cite{Winther2020}, and training of paramedics~\cite{schild2018epicsave}.
For a comprehensive review of past research on learning and transfer of complex motor skills in VR, please refer to~\cite{Levac2019}.
In the context of Human-Computer Interaction (HCI) research, evaluating and interacting with physical interfaces in VR~\cite{Paneva2020}, and conducting virtual field studies~\cite{Maekelae2020} have also been investigated. 
However, despite this extensive research, there are no studies focused specifically on the transferability of building with building blocks between physical and virtual environments.

\textit{Virtual Building Blocks for Learning and Creativity.}
In the realm of educational and training applications, building blocks have been integrated into virtual and mixed reality settings to enhance learning experiences and foster creativity.
For example, Zhao et al.~\cite{Zhao2019} developed a simulation game aimed at educating engineering students in manufacturing tasks, by designing and building LEGO cars in VR. 
The VR design tool Dreamscape Bricks VR~\cite{Doma2022dreamscape} utilises LEGO bricks as base components for creating architectural designs.
Another educational application, ScalebridgeVR~\cite{Pietroszek2019}, employs LEGO bricks to teach children proportional reasoning skills by challenging them to use to bricks to balance structures like bridges in VR.
BlocklyVR~\cite{Hedlund2023} focuses on teaching programming skills through the use of virtual building blocks in a VR setting.
In the context of Mixed Reality (MR), MagiBricks~\cite{Stefanidi2023} explored the use of smart toy bricks for distributed intergenerational play.
However, these studies primarily focused on the evaluation of the VR/MR building applications, and not on a comparison between the physical and virtual experience.

\textit{Summary.} Previous research has explored the transferability of skills across various tasks between physical and virtual environments, as well as the use of virtual building blocks to enhance learning and creativity. 
However, a gap remains in understanding how the experience of building with blocks translates between physical and virtual realms, which this study addresses by providing a structured comparison of LEGO building in both physical and VR settings.

\section{Apparatus}
For the LEGO building application in VR, we employed Unity as the IDE and the HTC Vive Pro with the SteamVR plugin as the VR hardware.
Users interacted with the application using the Vive handheld controllers.
The majority of assets, including the room and the interior decorations, were sourced from the Unity asset store. 
We converted 3D-printable models of a 16x16 baseplate and LEGO bricks of various sizes into a Unity-compatible format using Blender. 

\section{Pilot Study}
We conducted a pilot study with 4 participants (3 male, 1 female), aged between 22 and 26 years (mean 24.5, SD 1.91), to test the LEGO building VR counterpart and collect feedback on what (if any) additional features facilitated by the virtual environment users would like to have. 

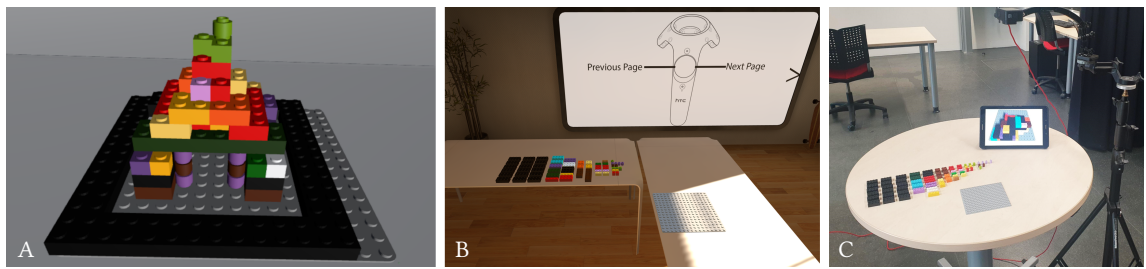
\begin{figure}[b]
    \centering
    \begin{tikzpicture}
    \draw (0, 0) node[inner sep=0] {    \includegraphics[width=\linewidth]{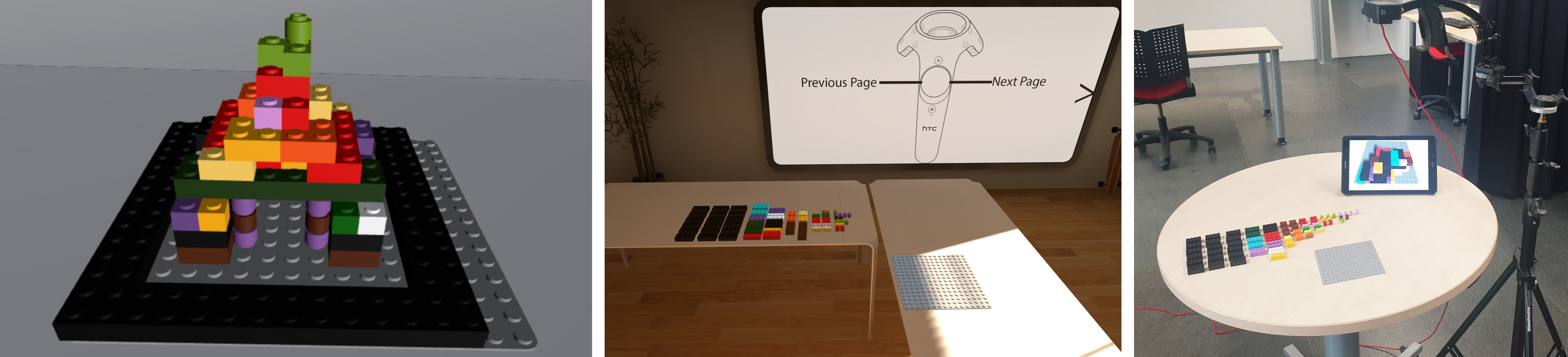}};
    \draw (-7.3, -1.5) node {\textcolor{white}{A}};
    \draw (-1.5, -1.5) node {\textcolor{white}{B}};
    \draw (3.6, -1.5) node {\textcolor{white}{C}};
    \end{tikzpicture}
    \caption{ In the pilot study, participants built the structure (A) in VR (B) and the real world (C).
    }
    \label{fig:PilotSetups}
\end{figure}

\textit{Setup and Procedure}
Participants were tasked with building the structure shown in \autoref{fig:PilotSetups}A, in two environments: the VR application (\textit{VR} condition) and the real world using physical LEGO (\textit{RW} condition). 
We aimed for a structure of medium difficulty that would present the right amount of challenge for the participants.
We validated that the structure is neither too easy nor too difficult in an informal test run before the pilot. 
In the \textit{VR} condition, participants found themselves in a furnished virtual room with directional lighting, where bricks and the baseplate were positioned on two tables (\autoref{fig:PilotSetups}B). 
Assembly instructions were presented on a wall-mounted virtual screen.
To maintain consistency, the layout in the \textit{RW} condition matched the VR setup (see \autoref{fig:PilotSetups}C). 
In the \textit{VR} condition, participants’ actions were screen-recorded using OBS\footnote{www.obsproject.com} for subsequent analysis, which included assembly time, and number and type of errors.
In the \textit{RW} condition, a tripod-mounted camera captured the real-world assembly (for both consent was obtained before the start of the experiment).
In a semi-structured interview, we collected data regarding the user experience in each environment, as well as suggestions for additional VR features.

\textit{Pilot Results}
Participants took an average of 20 minutes (SD 8) to complete the structure in \textit{VR}, compared to 8 minutes (SD 7) in the \textit{RW} condition. 
In these sessions, we observed several limitations in the VR interaction (addressed below), leading to an average of 46 (SD 26) dropped and 23 (SD 16) misplaced bricks in VR, compared to 1 (SD 1) dropped and 7 (SD 11) misplaced in \textit{RW}. 
Despite the longer assembly times in the \textit{VR} condition, participants expressed enthusiasm in the qualitative feedback, stating that they found the virtual environment pleasant, easy to focus on, and distraction-free.
Participants also appreciated the many options and creative possibilities afforded by the VR environment.

\textit{Prototype Iteration}
Unity's physics engine caused occasional glitches during brick manipulation, resulting in unexpected movements and placement errors. 
The bulky hand and controller models often led to unintended grabbing of bricks.
Additionally, the snapping system which required brick alignment (within some tolerance), was often not discovered by participants, causing confusion and delays.
To address these issues, we excluded held bricks from Unity’s physics simulations, replaced bulky hand models with simple physics-free high-accuracy spherical VR cursors, and improved the snapping mechanic to automatically snap bricks into place without requiring precise alignment.

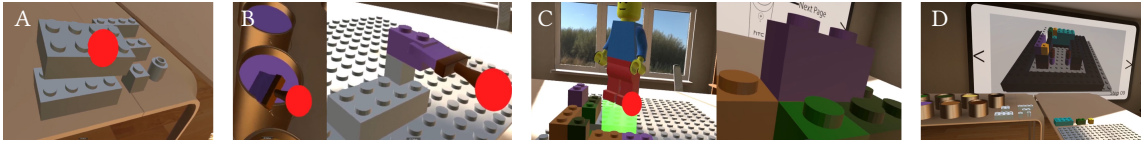
\begin{figure}[t]
    \centering
     \begin{tikzpicture}
    \draw (0, 0) node[inner sep=0] {    \includegraphics[width=\linewidth]{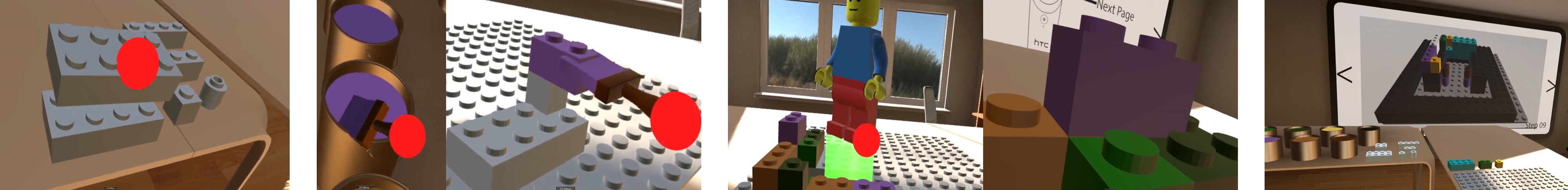}};
    \draw (-7.3, 0.7) node {\textcolor{white}{A}};
    \draw (-4.3, 0.7) node {\textcolor{white}{B}};
    \draw (-0.35, 0.7) node {\textcolor{white}{C}};
    \draw (4.9, 0.7) node {\textcolor{white}{D}};
    \end{tikzpicture}
    \caption{Superpowers for building with LEGO in VR. (A) Spawning a new brick from a starter brick. (B) Colouring a brick by using a virtual paintbrush and paint buckets. (C) Positioning and exploring the structure from the perspective of the LEGO figure. (D) Automatically spawning the required bricks for each step as the user turns the pages of the instructions.}
	\label{fig:superpowers}
\end{figure}

\textit{Additional Features}
The most frequently suggested additional features were aimed at removing some of the constraints of the physical world.
Specifically, participants expressed a keen interest in an infinite brick supply for unrestricted building, the ability to dynamically change brick colours, and the interesting capability to shrink themselves to LEGO figure size for immersive exploration of their structures.
A recurring source of frustration identified by participants was the 'tedious' process of searching for specific bricks.
To address this, some suggested adding a feature that automates the provision of the next required brick. 
We used these insights to inform the design of the following \textit{superpowers}:
\textbf{unlimited brick supply}, where we introduced starter bricks that serve as spawners for more bricks of the same type, providing users with virtually unlimited access to building materials (\autoref{fig:superpowers}A);
\textbf{colourable bricks}, adopting a painting metaphor that allows users to apply selected colours to LEGO bricks using a virtual brush and paint buckets (\autoref{fig:superpowers}B);
\textbf{shrinkable user perspective}, enabling users to resize themselves to the scale of a LEGO figure for exploring the structure from the minifigure’s viewpoint (\autoref{fig:superpowers}C); and 
\textbf{automatic brick provision}, where the required bricks for the next step are automatically spawned as users turn the pages of the building instructions, reducing search time, ensuring needed pieces are always within reach, and preventing clutter (\autoref{fig:superpowers}D).
Building on this insight, we improved the VR-counterpart application and implemented the additional features into a separate version.

\section{User Study}

In the user study, we compared user experience, assembly efficiency, and overall engagement across two VR applications (with and without \textit{superpowers}) and real physical LEGO. Our goal was to gain a deeper understanding of how effective and appealing the VR environments are in both structured and creative building scenarios, and to gather insights into how the additional VR features influence the building process.

\textit{Experimental Design}
The user study followed a within-groups experimental design.
In addition to the \RW and \VR conditions, we now include the VR with additional features in the evaluation (\AVR condition).
The experimental task included both an assembly task and time for creative free building.
All conditions were presented in a randomised order.

\textit{Participants}
22 participants (8 female, 14 male), aged between 23 and 62 years (mean 29.45, SD 10.56) with normal or corrected-to-normal vision, participated in the study. 
Before the experiment, participants read basic information about the study and signed a consent form. 
All participants received a monetary reimbursement for their participation.

\textit{Procedure and Measures}
Participants were asked to take a comfortable position, either standing up or sitting down, that allowed for a good overview of all the LEGO bricks and the baseplate. 
In the \VR and \AVR conditions, participants were asked to adjust the HTC Vive Pro headset to their comfort, and were provided with the Vive controllers.
Similar to the pilot study, for the assembly task, participants were instructed to follow the provided instructions and recreate the given structure as fast and accurately as possible.
For the creative free building phase of the experiment, participants were asked to build a structure of their choice in each condition.
We prescribed the general topic of architecture for easier comparison of the results. 
We recorded the time, and if the creative building exceeded 5 minutes, we asked participants to move to the next condition. 
Following each building session, a questionnaire containing study and condition-specific questions was administered.  
Upon completing the assembly and the creative task in all there conditions, amounting to six sessions in total, a semi-structured interview was conducted.
The approximate duration of the study was 60 minutes.

\section{Results}
\label{sec:Results}

\subsection{User Experience}
\label{subsec:User_Experience}

\subsubsection{Perceived Fun}
\label{ssection:Perceived_Fun}

Participants rated the level of fun they experienced during the tasks on a scale from 1 (not fun at all) to 10 (extremely fun).
The average reported level of fun for the assembly task was 8.09 (SD 1.48) for the \textit{RW}, 8.73 (SD 1.83) for the \textit{VR}, and 8.45 (SD 1.84) for the \textit{AVR} condition. 
Using the Friedman test, we observed a significant difference between the conditions (p=0.0036). 
However, this difference did not survive the post-hoc pairwise test (p>0.05).
For the free building task, the average perceived fun was the highest for the \textit{AVR} condition (mean 9.32, SD 0.99), followed by \textit{VR} (mean 8.91, SD 1.11) and \textit{RW} (mean 8.09, SD 2.11).
Again, the significant difference between the conditions (p=0.011) did not survive post-hoc testing (p>0.05).



\subsubsection{Preferred Conditions}
\label{par:PreferredCondition}

When participants were asked which environment they preferred for the assembly task, 10 participants chose the \AVR condition, 9 preferred the \RW condition, and only 3 chose the \VR condition. 
Some reasons provided for choosing \textit{AVR} included the many possibilities, customisation options, and the overall impression of being more fun.
Participants who opted for the \textit{RW} condition appreciated the haptic feedback, intuitive brick manipulation, the absence of depth cue issues, and the tangible outcome that can be held and observed later.
Moreover, they found it easier to build under time pressure in the \textit{RW} condition due to their familiarity and past experience with real LEGO. 
Meanwhile, the 3 participants that chose the \textit{VR} condition said that they found it to be more fun than \textit{RW}, while not being as 'easy' as \textit{AVR}.
For the free creative building task, the overwhelming majority of participants (20) chose \AVR as their preferred condition, only 2 participants chose the \RW condition, and none chose \VR. 
In the subsequent interviews, among the 20 participants that opted for the \textit{AVR} condition, 7 explicitly mentioned the LEGO figure's perspective superpower as a reason they chose this condition.
Additionally, 2 participants appreciated the ability to construct structures in AVR that would be challenging or even impossible with real LEGO as they would become unstable or fall apart under the influence of gravity. 
Examples of these structures are shown in \autoref{fig:time}C. 
Participants also provided more general reasons for their preference of the \textit{AVR} condition, such as it being more interesting and fun, and allowing for many building possibilities due to its customisation features.
Out of the 2 participants choosing the \textit{RW} condition, one person noted the lack of haptic feedback in the VR conditions as the main reason, while the other expressed a dislike for the fact that the VR conditions didn't require as much fine motor skills as the \textit{RW} condition.

\subsubsection{Enjoyment Of Additional Features}
\label{par:EnjoymentOfAdditionalFeatures}

After completing the assembly in the \AVR condition, we asked participants to rate how much they enjoyed the different superpowers on a scale from 1 (did not enjoy it at all) to 10 (enjoyed it a lot).
Overall, participants rated the features highly: automatic brick provision received 8.14 (SD 2.59), unlimited brick supply 9.45 (SD 1.92), colourable bricks 8.86 (SD 2.01), and the ability to shrink to a LEGO figure’s perspective 9.41 (SD 1.59).
These scores are consistent with the qualitative feedback, with most participants expressing a favourable view towards the additional features. 
Nevertheless, there were divergent opinions, particularly concerning the automatic brick provision. 
While it simplifies the task, participants who gave lower scores to this superpower argued that the process of searching for bricks is part of the challenge, contributing to the enjoyment and the sense of accomplishment.


\subsection{Performance}
\label{subsec:Performance}

\subsubsection{Assembly Completion Times}
\label{subsubsec:AssemblyTimes}

\begin{figure}[b]
  \centering
      \begin{tikzpicture}
        \draw (0, 0) node[inner sep=0] {
        \includegraphics[width=0.95\linewidth]{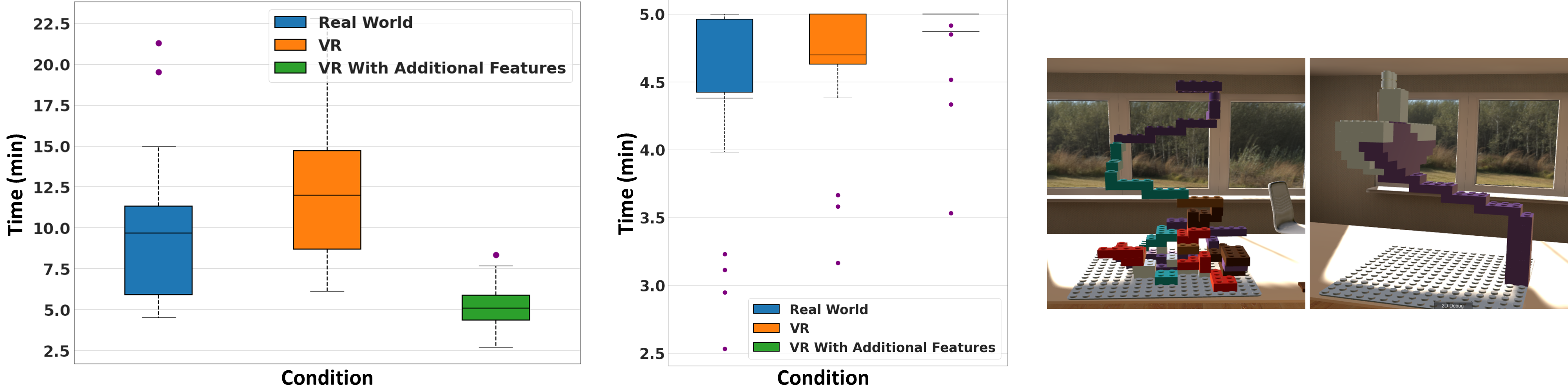}};
        \draw (-6.25, -1.35) node {\textcolor{black}{A}};
        \draw (-0.85, -1.35) node {\textcolor{black}{B}};
       \draw (2.65, 1) node {\textcolor{white}{C}};
       \draw (4.35, 1) node {\textcolor{white}{\textit{AVR}}};
        \draw (6.8, 1) node {\textcolor{white}{\textit{AVR}}};
    \end{tikzpicture}
\caption{Time taken for the assembly task (A) and free creative building (B) across the \textit{RW}, \textit{VR}, and \textit{AVR} conditions. (C) Creative structures built by participants ID 3 (left) and ID 4 (right) in \textit{AVR}, which would be challenging to construct in the physical world.}
\label{fig:time}
\end{figure}
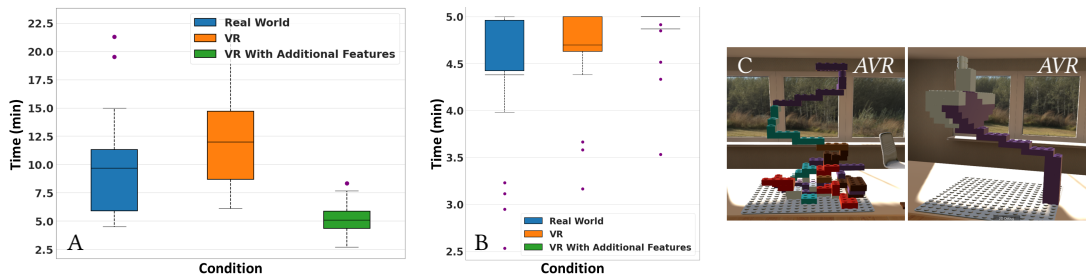

The completion times for the assembly task are shown in \autoref{fig:time}A.
On average, participants needed the most time in the \VR condition (mean 11.97, SD 4.43), followed by the \RW condition (mean 9.67, SD 4.57), while they completed the assemblies in the \AVR condition the fastest (mean 5.07, SD 1.32).
The Friedman test indicated a significant difference between the conditions (p<0.001).
A post-hoc test using Wilcoxon tests with false discovery rate correction showed significant differences between the \RW and \AVR (Wilcoxon p<0.001), as well as the \VR and \AVR (Wilcoxon p<0.001) conditions.

\subsubsection{Assembly Errors}
\label{par:AssemblyMistakes}

The number of errors (incorrect brick type or placement) that occurred during the assemblies was similar across \RW (mean 11.41, SD 8.40), \VR (mean 13.82, SD 11.10), and \AVR (mean 14.18, SD 8.82).
A Friedman test showed no statistical significance between the conditions (p=0.28).
We observe that the automatic brick provision superpower in \textit{AVR} did not seem to impact error rates, suggesting that errors were more likely due to incorrect placement rather than using the wrong brick type.

\begin{figure}[t]
    \centering
    \begin{tikzpicture}
        \draw (0, 0) node[inner sep=0] {
        \includegraphics[width=.26\linewidth]{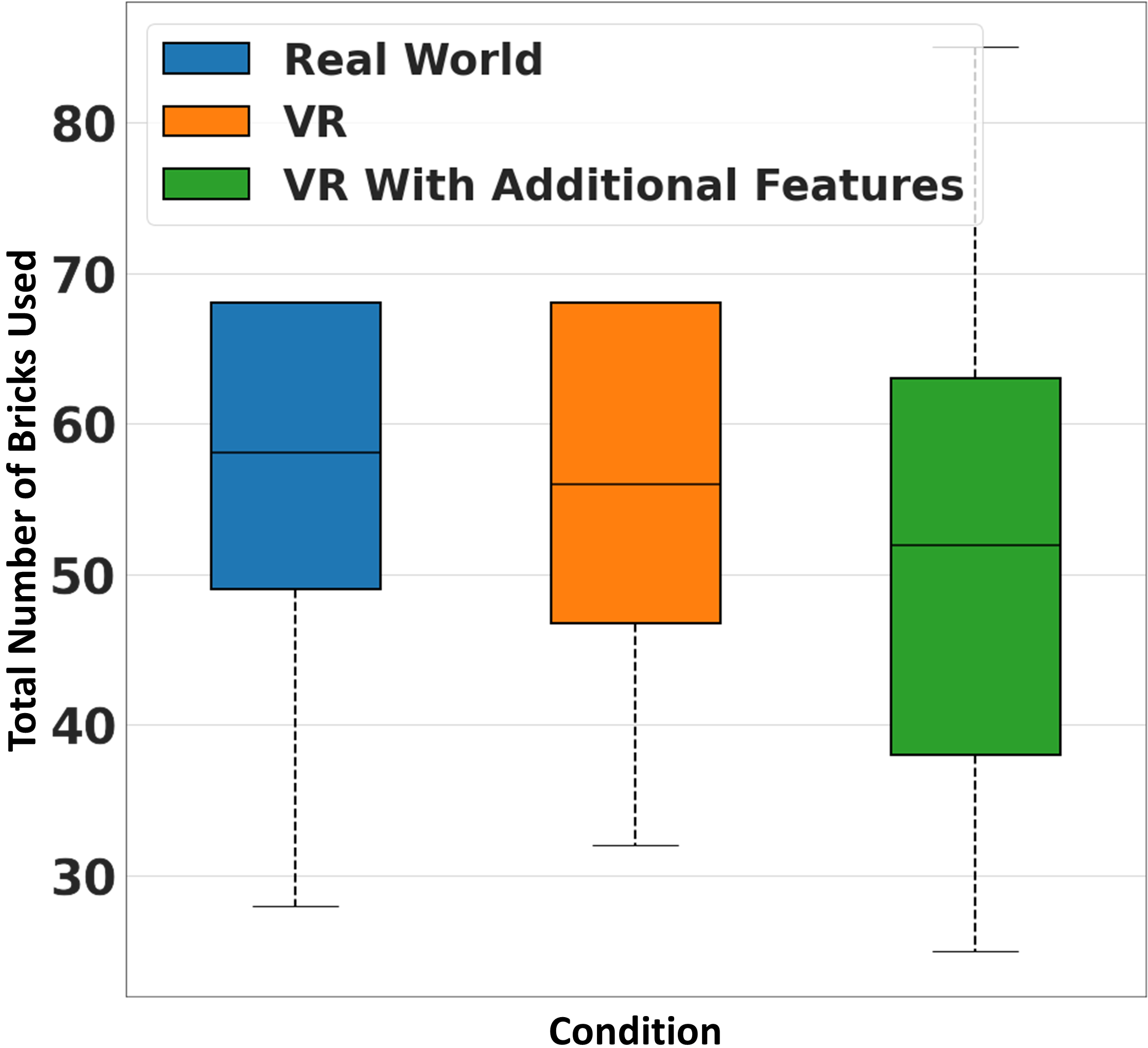}
        \includegraphics[width=.26\linewidth]{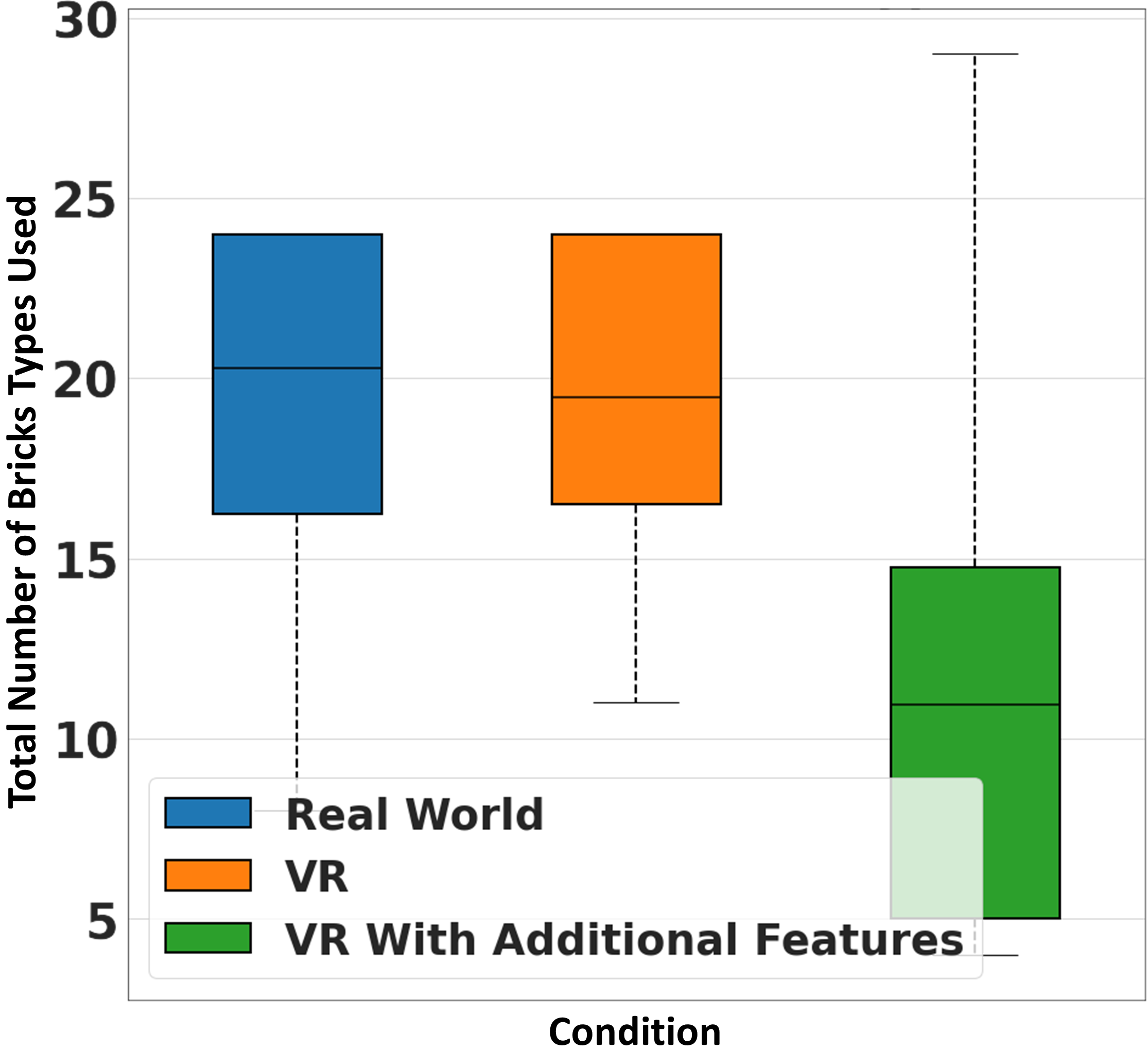}
        \includegraphics[width=.46\linewidth]{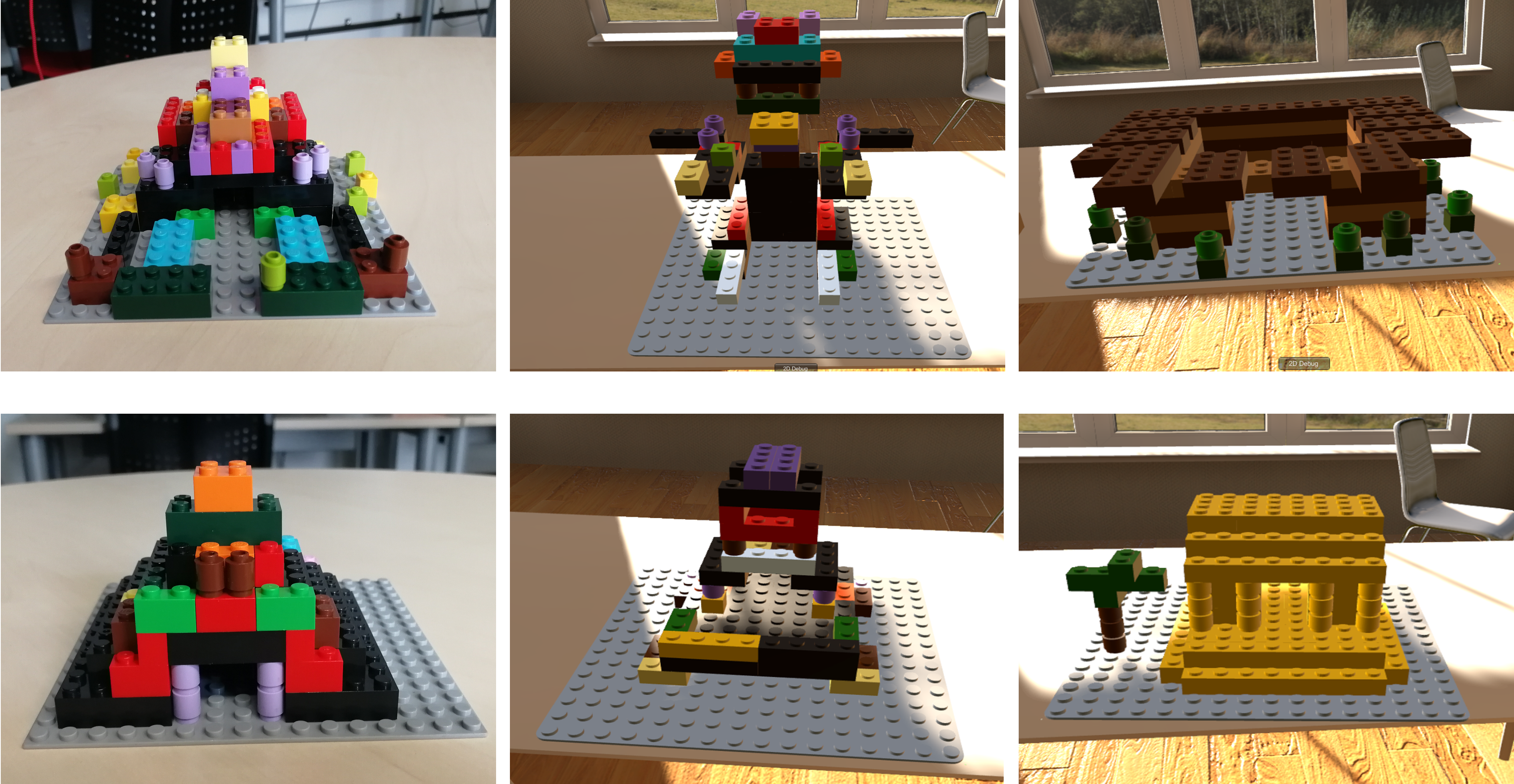}};
        \draw (-6.85, -1.45) node {\textcolor{black}{A}};
        \draw (-2.85, 1.6) node {\textcolor{black}{B}};
       \draw (0.7, 1.6) node {\textcolor{white}{C}};
        \draw (2.5, -0.25) node {\textcolor{white}{\textit{RW}}};
        \draw (2.5, 1.6) node {\textcolor{white}{\textit{RW}}};
        \draw (4.85, -0.25) node {\textcolor{white}{ \textit{VR}}};
        \draw (4.85, 1.6) node {\textcolor{white}{ \textit{VR}}};
        \draw (7.15, -0.25) node {\textcolor{white}{ \textit{AVR}}};
        \draw (7.15, 1.6) node {\textcolor{white}{ \textit{AVR}}};
    \end{tikzpicture}
    \caption{ Comparison of the total number of bricks (A) and the variety of brick types (B) used across the \textit{RW}, \textit{VR}, and \textit{AVR} conditions. (C) Comparison of the creative structures built by two participants (top row ID 19, bottom row ID 22) in each of the three conditions. 
    }
    \label{fig:BrickUsage}
\end{figure}
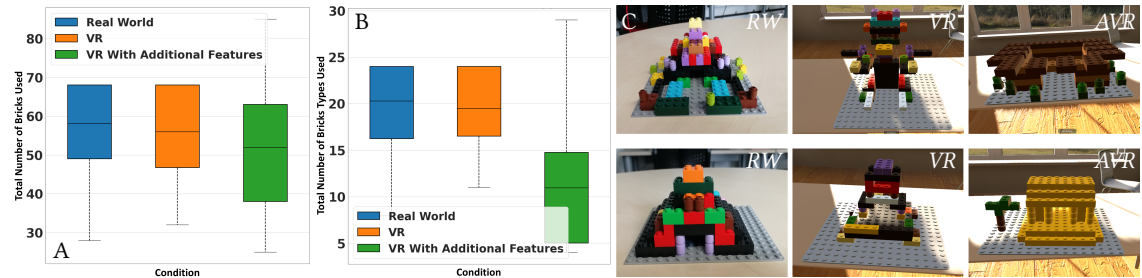

\subsection{Duration and Brick Usage in the Creative Task}
\label{subsec:Duration_BrickUsage}

\subsubsection{Duration}
\label{par:DurationFreelyBuilding}

On average participants spent 4.38 minutes (SD 0.74) in the \RW condition, 4.69 minutes (SD 0.53) in \textit{VR}, and 4.87 minutes (SD 0.34) in \textit{AVR} (see \autoref{fig:time}B).
A Friedman test indicated a significant difference between the conditions (p<0.001).
The Wilcoxon test with false discovery rate correction showed that participants spent significantly more time freely building in the \AVR compared to the \RW condition (p=0.0029).

\subsubsection{Brick Usage}
\label{par:brickUsage}

\autoref{fig:BrickUsage} shows the total number of LEGO bricks used during the creative free building task.
Participants used an average of 51.91 (SD 16.14) bricks in \RW, 56.0 (SD 12.23) in  \VR, and 58.05 (SD 0.48) in \AVR.
A Friedman test did not indicate a significant difference between the conditions (p=0.11).
We also analysed the usage of different brick types, where each distinct shape and colour combination counts as a separate type. 
On average, participants used the most brick types in \RW (mean 20.27, SD 4.93), followed by \VR (mean 19.45, SD 4.19), and the least in \AVR (mean 10.95, SD 7.18), as shown in \autoref{fig:BrickUsage}B.
A Friedman test indicated a significant difference between the conditions (p<0.001).
The difference in the total number of brick types used was statistically significant between the \RW and \AVR (Wilcoxon p<0.001), as well as between the \VR and \AVR conditions (Wilcoxon p<0.001).
Structures built in \textit{AVR} tended to have a more uniform monochromatic appearance, with prominent examples shown in \autoref{fig:BrickUsage}C. 

\section{Discussion}
\label{sec:Discussion}


\textit{Performance and User Experience.}
Participants completed the assembly task significantly faster in \textit{AVR}, compared to both \textit{RW} and \textit{VR}.
This improvement in performance is most likely due to the elimination of search time for needed bricks, as bricks were automatically provided.
However, despite the faster completion times, the number of errors remained comparable across all conditions. 
This suggests that errors were primarily due to the incorrect placement of bricks rather than the selection of the wrong bricks, which the automatic brick provision feature in \textit{AVR} effectively addressed.
When asked about their preferred conditions, participants were almost equally split between \textit{AVR} and \textit{RW} for assembly tasks but overwhelmingly chose \textit{AVR} for creative tasks.  
Some participants stated that they preferred \textit{RW} for tasks where the focus is on efficiency, due to the familiarity and tactile feedback of real LEGO.
In contrasts, a frequently named reason for preferring \textit{AVR} for creative tasks was the ability to shrink to the perspective of a LEGO figure and explore structures from within, a feature perceived to enhance creative engagement by uncovering new aspects of the building experience.

\textit{Creativity and Engagement.}
When it came to free creative building, on average participants spent the most time in \textit{AVR}, significantly more than in \textit{RW}.
This extended engagement time in \textit{AVR} suggests that while the environment offered enhanced capabilities, it also presented a level of complexity that required users to spend more time planning and conceptualising their builds. 
Some participants reported feeling overwhelmed by the many options available in \textit{AVR}, needing additional time to decide on what they wanted to build, whereas in \textit{RW} and \textit{VR}, they immediately started building with the bricks at hand.
Interestingly, despite the many options in \textit{AVR}, many participants opted to use a less diverse set of bricks, resulting in less colourful and more uniform-looking structures.


\textit{Implications for VR Design.}
The faster completion times observed in \textit{AVR}, due to the automatic provision of bricks, suggest that automation can significantly improve task efficiency. 
However, user feedback indicated that the ease introduced by automation may diminish some of the challenge that makes building with LEGO fun and engaging. 
For VR designers, this highlights the importance of \textbf{balancing automation with user control}. While automation can reduce the cognitive load associated with repetitive tasks, it is crucial to maintain a level of user agency to ensure that the experience remains engaging and meaningful.
The extended time participants spent on the creative task in \textit{AVR}, coupled with some reporting initial feelings of being overwhelmed by the many options, presents a design challenge: how to provide users with ample \textbf{creative freedom without overwhelming them}. 
VR designers should consider implementing intuitive filtering or recommendation systems that help users navigate large sets of options, and only gradually introduce complexity as users become more proficient. For example, frameworks like \textit{SelVReflect}~\cite{Wagener2023} emphasize progressive guidance and adaptive interfaces, offering insight for structuring and scaffolding creative open-ended tasks in VR. 

\textit{Limitations and Future Research.}
In our study, we used controllers as the input method for the VR applications, reflecting the current consumer-grade standard for VR interactions and providing a widely accessible and consistent interface for users.
However, future research could explore the impact of different input methods on the LEGO building experience in VR, such as gaze and pinch interaction seen in recent systems like the Apple Vision Pro, or hands-only techniques used in the Hololens 2 or the Meta Quest series. 
While these methods may not improve the quality of the haptic feedback when manipulating the building blocks, they could offer a more intuitive interaction experience.
As input technologies continue to evolve, performing building block tasks in VR could offer new opportunities for guiding and facilitating the development of creative fine motor skills, particularly in educational, training, and rehabilitation contexts. 
By enabling the adjustment of different task parameters and environmental conditions on-the-fly, as well as offering continuous progress monitoring and feedback, VR could potentially become a powerful tool for enhancing the effectiveness of various learning and skill development interventions.

\section{Conclusion}
\label{Conclusion}
The paper presented a structured comparison of the LEGO building experience across three environments: the physical world, VR, and VR enhanced with additional features termed as \textit{superpowers}. 
Through a comprehensive study involving structured assembly and creative free-building tasks, we analysed differences in user experience, performance, and engagement across these environments.
Our findings reveal that while the physical environment continues to offer a familiar tactile experience that is particularly valued in assembly tasks, the VR superpowers present interesting opportunities for creativity and engagement, particularly in free-building scenarios. 
However, the added complexity of the virtual setting can be both a source of innovation and a challenge for users, highlighting the need for careful design to balance creative freedom with usability.
Future research should continue to explore the potential of VR technologies, especially as input methods and haptic feedback mechanisms evolve, to further bridge the gap between physical and virtual building experiences, with particular relevance for educational, training, entertainment, and therapeutic contexts.



\begin{acks}
The authors thank all the participants of the pilot and the user study. 
\end{acks}

\bibliographystyle{ACM-Reference-Format}
\bibliography{sample-base}




\end{document}